# ON THE THREE LAWS OF EARTHQUAKE PHYSICS

A.V. Guglielmi, A.D. Zavyalov, O.D. Zotov, B.I. Klain

*Schmidt Institute of Physics of the Earth, Russian Academy of Sciences; Bol'shaya Gruzinskaya str., 10, bld. 1, Moscow, 123242 Russia;*
*e-mail: guglielmi@mail.ru, zavyalov@ifz.ru,*
*ozotov@inbox.ru, klb314@mail.ru*

The paper provides a synoptic overview of a series of works carried out by a group of researchers at the Institute of Physics of the Earth RAS with the aim of finding new approaches to the problems of earthquake physics. The fundamental laws of Omori, Gutenberg-Richter and Bath have served as a constant support and reference point in the course of many years of research. The concept of the tectonic earthquake triad as a natural trinity of foreshocks, main shocks and aftershocks is used in the article to organise the thematic material. A classification of main shocks into six types of triads found in experience is given. The parameters appearing in the three laws for different types of triads are given. The axiomatic theory of the evolution of aftershocks is outlined. The concepts of source deactivation, Omori epoch and source bifurcation are introduced, and the notion of the proper time of unsteady lithospheric processes is introduced. Convergence of foreshocks and divergence of aftershocks are mentioned. The general conclusion is that the Omori, Gutenberg-Richter and Bath laws are reliable tools in the experimental and theoretical study of earthquakes. The laws have a depth of content that has been demonstrated by the ability to enrich the original formulations of the discoverers with interesting and important additional statements.

The paper is submitted to the journal "Volcanology and Seismology".

## 1. Introduction

Three empirical laws of earthquake physics are widely known and actively used in research: the Omori law, the Gutenberg-Richter law, and the Bath law. Chronologically, the first to be established was Omori's law, which states that after the main shock of a tectonic earthquake, the frequency of repeated tremors (aftershocks) decreases hyperbolically over time. Mathematically, the law is expressed by the formula

$$n(t) = \frac{k}{c+t}, \tag{1}$$

where $n$ is the average frequency of aftershocks, $k > 0$, $c > 0$, $t \geq 0$ [Omori, 1894].

The Gutenberg-Richter law expresses the exponential dependence of the number of earthquakes $N$ on the magnitude $M$:

$$N = 10^{a-bM}. \qquad (2)$$

Parameters $a$ and $b$ characterize the seismic activity of a given region over a certain time interval [Gutenberg, Richter, 1944].

Finally, Bath's law states that the difference $\Delta M = M_0 - M_{max}$ between the magnitude of the main shock $M_0$ and the maximum magnitude $M_{max}$ in the aftershock sequence, firstly, does not exceed

$$\Delta M = 1.2, \qquad (3)$$

and, secondly, does not depend on $M_0$ [Bath, 1965].

All three laws are statistical. They are probabilistic in nature, relate to average values and are recognized by the scientific community as fundamental statements expressing in brief form stable objectively existing connections and relationships (see, for example, monographs [Richter, 1963; Bolt, 1981; Kasahara, 1985; Mogi, 1988]). The laws are successfully used in seismology to analyze and systematize the facts discovered during the observation of earthquakes. Not being a consequence of deductive theory, the laws themselves serve as a reliable basis for forming hypotheses, for constructing phenomenological theories and for developing mathematical models of earthquakes.

At the same time, with the development of earthquake physics, the laws are enriched with interesting details, important clarifications, and sometimes the laws are subject to radical revision. This is natural and inevitable, and is entirely consistent with the Cartesian concept of the methodology of scientific research [Descartes, 1953]. Let us give one example. In 1924, that is, 30 years after the formulation of Omori's law, Hirano formulated his version of the law of aftershock evolution [Hirano, 1924]. He proposed replacing (1) with formula

$$n(t) = \frac{k}{(c+t)^p} \qquad (4)$$

on the grounds that the power-law dependence better approximates the observed aftershock data than the hyperbolic one. Here $p$ is an additional parameter, $p > 0$ (see also [Jeffreys, 1938; Utsu, 1961, 1962; Ogata, 1988; Utsu et al., 1995; Ogata, Zhuang, 2006; Rodrigo, 2021; Salinas-Martínez et al., 2023]).

This paper presents a synoptic review of our experience in the long-term search for new approaches to the problems of earthquake physics (see papers [Guglielmi, 2015, 2016, 2017; Zavyalov et al., 2020, 2022; Guglielmi et al., 2023] and the literature cited therein). The fundamental laws of Omori, Gutenberg-Richter and Bath have served as constant support and guide for us in our research. Below we will give an idea of the diversity of earthquake triads, the deactivation of the source, the proper time of the earthquake source, the Omori epoch and the mirror analogue of Bath's law. We will also discuss questions about the status of laws and hypotheses, about phenomenology and about the axiomatic method.

## 2. Main shock
### 2.1. Classification of main shocks

We will call the natural trinity of foreshocks, main shock and aftershocks an earthquake triad [Guglielmi, 2015]. Let $N_-$ be the number of foreshocks that occurred during a certain period of time preceding the mainshock, and $N_+$ be the number of aftershocks during a period of the same duration after the mainshock. Let us choose the following basis for dividing triads into classes. If $N_- < N_+$, then we will call the triad classical. This class is widely known and well studied. It is known that in many cases foreshocks are absent, i.e. $N_- = 0$. If $N_- = 0$, then we will call the triad incomplete or shortened, but if $N_- \neq 0$, then we will call the triad complete. Thus, we have a feature for dividing classical triads into two types. Figure 1 shows the classes and types of triads.

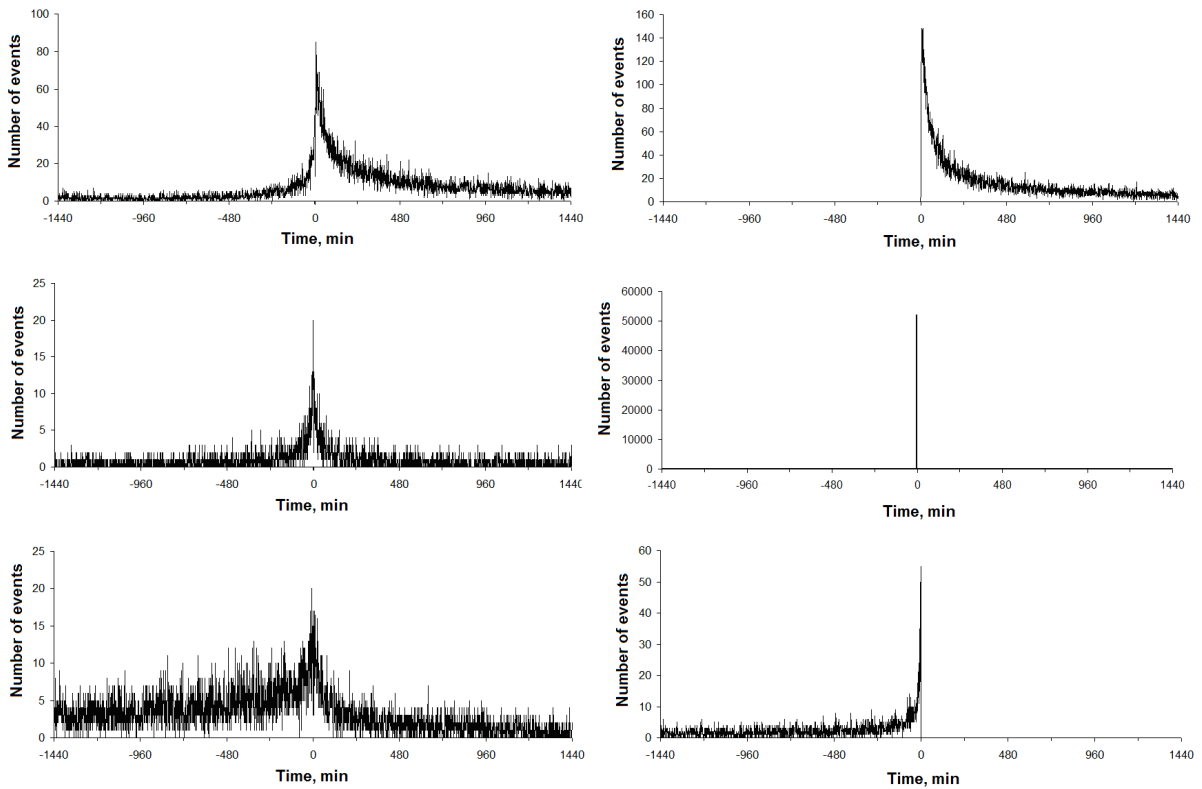

**Fig. 1**. Three classes and six types of triads. From top to bottom: classical, symmetrical and mirror triads. From left to right: complete and incomplete triads. The zero point on the time axis corresponds to the times of the main shocks.

The graphs were obtained by synchronous summation of triads for the period from 1973 to 2019 based on data from the USGS/NEIC catalog. The magnitudes of the main impacts were $M_0 = 5-8$, the depths of their hypocenters were 0 – 250 km.

Thus, we have three classes and six types of triads. Foreshocks and/or aftershocks may be absent, but the main shock is an essential attribute of each class and each type. This gives us the opportunity to classify the main shocks according to their belonging to one or another class and type of triad. The types of main shocks can be divided into varieties according to one or another characteristic, for example, according to the depth of the hypocenter, according to the focal mechanism, taking into account, however, that there must be only one characteristic at a given level of division.

Let us introduce a rectangular matrix $T(M_0)$ for the abbreviated designation of classes and types of main impacts with a magnitude of $M \geq M_0$:

$$T(M_0) = \begin{pmatrix} TC1 & TC2 \\ TS1 & TS2 \\ TM1 & TM2 \end{pmatrix}. \qquad (5)$$

The letter $T$ stands for the word "Triad". (Recall that the classification of main shocks coincides with the classification of triads.) Similarly, $C$, $S$ and $M$ stand for the words "Classical", "Symmetrical" and "Mirror". The number 1 indicates that the triad is complete, and the number 2 indicates that the triad is shortened. The grounds on which triads are divided into classes and types are indicated in Table 1.

**Table 1.** Criteria for dividing triads into classes and species.

| Class | | | | | |
|---|---|---|---|---|---|
| TC | | TS | | TM | |
| $N_- < N_+$ | | $N_- = N_+$ | | $N_- > N_+$ | |
| Specie | | | | | |
| TC1 | TC2 | TS1 | TS2 | TM1 | TM2 |
| $N_- \neq 0$ | $N_- = 0$ | $N_\pm \neq 0$ | $N_\pm = 0$ | $N_+ \neq 0$ | $N_+ = 0$ |

The systematics we have developed opens up the possibility of asking new questions about the morphology and physics of main shocks. As a first step along this path, we raised the question of the probability of excitation one or another type of main shock. Figure 2 shows the results of a study of this issue based on data from 5600 main shocks with magnitudes $M \geq 6$.

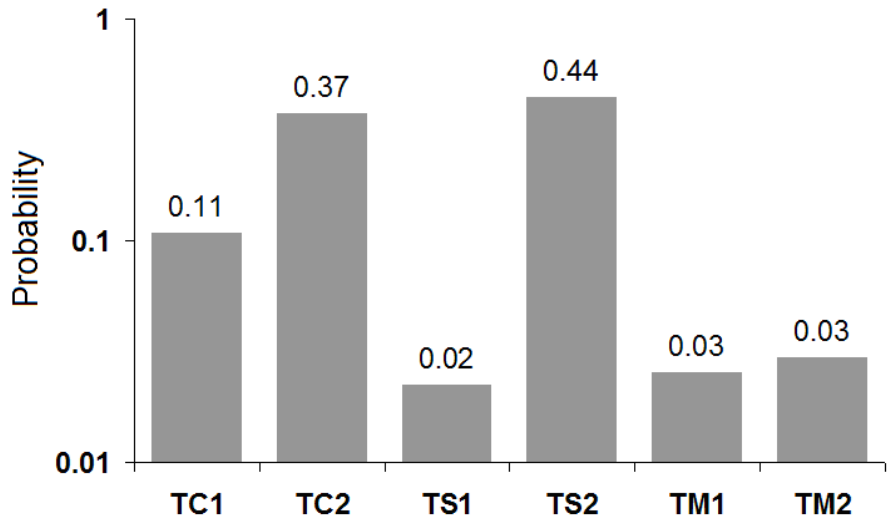

**Fig. 2**. Probability of observing main shocks with magnitudes $M \geq 6$, belonging to different types of triads.

Note that none of the six species is empty. Furthermore, in almost half of all cases, not only foreshocks are absent, but also aftershocks. Finally, it can be noted that the main beats in the shortened triads are several times more than in the full triads.

## 2.2. Gutenberg-Richter law

The parameter $a$ in the Gutenberg-Richter law (2) indicates the overall earthquake activity in the region. Figure 2 gives a qualitative idea of the parameter a for the main shocks that are part of various triads. In a certain sense, the parameter $b$ is more interesting. The smaller $b$, the higher the ratio of strong earthquakes to weaker ones. A typical value of $b = 1$, but generally speaking, $b$ varies widely, approximately from 0.5 to 1.5.

The question naturally arises about the distribution of the value of $b$ by types of main shocks in our classification system. We have presented this distribution in Figure 3. The drawing allows us to see previously unknown properties of the main shocks.

The first thing that catches the eye is the elevated $b$ values in the incomplete triads in each of the three classes. This difference is most strongly expressed in the class of symmetrical triads. It remains to be seen what factors influence the apparent difference in $b$ in full and shortened triads.

Furthermore, the lowest $b$ value is observed in the full classical TC1 triads. Finally, the highest b value is observed in incomplete symmetrical triads TS2. The abundance of triads of the TS2 type (see Figure 2) attracts our closest attention. In this regard, we have given the TS2 species the special name *Grande terremoto solitario* (abbreviated GTS). The point is this.

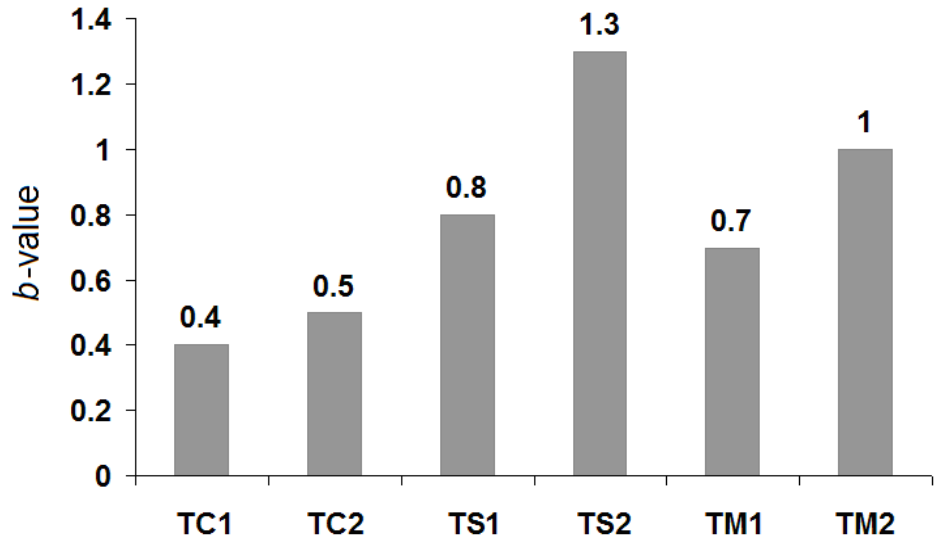

**Fig. 3**. The key parameter *b* of the recurrence graph for main shocks belonging to different types of triads.

We have already noted above the widespread opinion in the literature that the typical value of *b* is equal to 1. The value *b* = 1 is indicated in many well-known works [Mogi, 1962; Scholz, 1968; Mori, Abercombie, 1997; Schorlemmer et al., 2005]. But our analysis of the Gutenberg-Richter law shows that this view may be inaccurate. The value *b* = 1 is typical only for the extremely rare type of TM2 main impacts. In the vast majority of cases, the *b* value is either significantly lower than the specified value (TC class main shocks) or significantly higher (GTS). In other words, the value *b* = 1 is not typical for main shocks.

### 2.3. Bath's law

In connection with our discovery of a variety of main shocks, the problem of refining Bath's law for those types that are accompanied by aftershocks arose. These are the main shocks in the full and shortened classical triads, the full mirror triad and the full symmetrical triad. In this work we will solve the problem only partially. We will analyze the gap $\Delta M$ using the linear relation

$$\Delta M = aM_0 + b \qquad (6)$$

for the classical triads, full and shortened together, and for the full mirror triad. Here *a* and *b* are fitting parameters.

The data for the analysis were taken from the USGS catalog. In the interval from 1973 to 2019, main shocks with magnitudes $M_0 = 5-8$ and hypocenters at depths of 0–250 km were selected. Aftershocks were selected at intervals of 24 hours after the main shock in epicentral zones, the radius of which was determined using the methodology described in the work [Zavyalov, Zotov, 2021]. The analysis was carried out using the method of superimposing epochs. The result of the analysis of classical triads is presented in Figure 4. The solid line in the lower panel approximates the experimental points. The coefficients in formula (6) are $a = 0.48$, $b = -1.85$. We found that, contrary to popular belief, the magnitude of $\Delta M$ depends on $M_0$ for classical triads. For complete mirror triads, a dependence of $\Delta M$ on $M_0$ is also observed, with $a = 0.44$, $b = -1.61$.

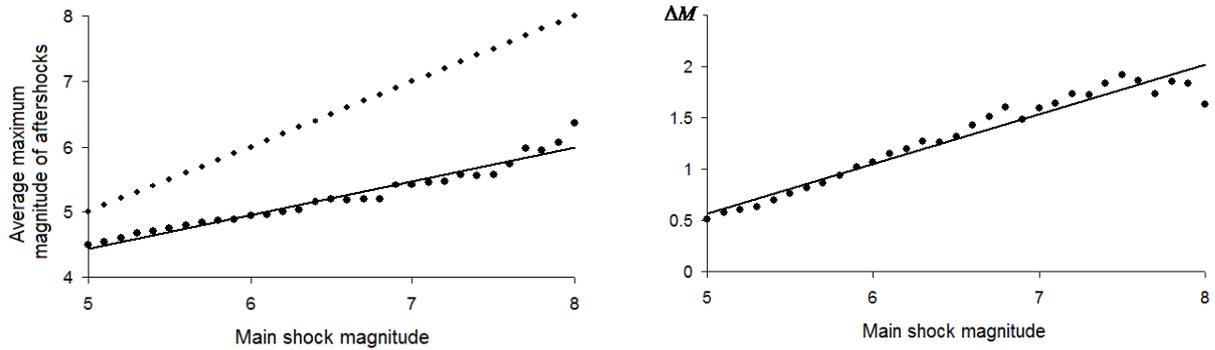

**Fig. 4**. Illustration of Bath's law for classical triads. The left panel shows the mainshock magnitudes and the average maximum aftershock magnitudes (upper and lower rows of dots, respectively). The right panel shows the gap $\Delta M$ as a function of the mainshock magnitude.

## 3. Aftershocks

### 3.1. Elementary theory of aftershocks

Let us present the elementary theory of aftershock evolution axiomatically. The earthquake source, the source deactivation coefficient $\sigma$, and the average aftershock frequency $n$ are the basic concepts of the theory. The source is a dynamic system, the current state of which is described by the deactivation coefficient. The deactivation coefficient is a continuous function of time. The average aftershock frequency is a substantially positive continuous smooth function of time. The theory is based on the following postulate: The deactivation coefficient is calculated using the formula

$$\sigma = -\frac{1}{n^2}\frac{dn}{dt} \qquad (7)$$

From (7) follows a statement (theorem), the content of which makes the choice of postulate obvious. Namely, the classical Omori law (1) is satisfied if and only if the deactivation coefficient is independent of time. Indeed, if $\sigma = \text{const}$, then (1) follows from (7). If, on the other hand, the aftershock frequency decreases hyperbolically with time, then $\sigma = \text{const}$, as follows from formula (7).

Let us present a number of other theorems that enrich the elementary theory. The evolution of aftershocks is described by a linear differential equation

$$\frac{dg}{dt} = \sigma(t), \tag{8}$$

where $g(t) = 1/n(t)$ is an auxiliary function. If the auxiliary function is proportional to time, then Omori's law (1) follows from (8). Further, for $\sigma = \mathrm{const}$, the Omori law (1) is the only solution to the nonlinear differential equation

$$\frac{dn}{dt} + \sigma n^2 = 0 \tag{9}$$

Finally, when $\sigma = \sigma(t)$, the general solution of the evolution equation (9) has the form

$$n(t) = \frac{n_0}{1 + n_0 \tau(t)},$$
$$\tau(t) = \int_0^t \sigma(t)\,dt \tag{10}$$

where $n_0 = n(0)$ is the initial condition. Formula (10) flexibly models the evolution of aftershocks and is a natural generalization of Omori's law [Guglielmi, 2016].

The deductive method of forming a theory allowed us to see the versatility of Omori's law. But this is not the only thing that has supplemented the aftershock theory. Using the quasilinear evolution equation (9), we can pose the inverse problem of a source "cooling down" after the main shock. Solving the inverse problem gives us an effective method for processing and analyzing aftershocks. So we want to calculate the deactivation factor from the observed aftershock frequency data. In practice, the original function $n(t)$ usually fluctuates rapidly. Therefore, the formal solution $\sigma = dg/dt$ turns out to be incorrect. Regularization can be conveniently performed by optimally smoothing the auxiliary function. The correct solution to the inverse problem has the form

$$\sigma(t) = \frac{d}{dt}\langle g(t) \rangle. \tag{11}$$

Here the angle brackets denote the smoothing procedure.

### 3.2. The Omori epoch

Two empirical laws of aftershock evolution, namely, the Omori law expressed by formula (1) and the Hirano-Utsu law expressed by formula (4), are intended to have predictive power. It is sufficient to select the initial condition $n(0) = n_0$ and calculate the phenomenological parameters at the initial stage of evolution. After this, based on the law, it is possible to predict with varying accuracy the activity of aftershocks at the subsequent stage of evolution. To use the closing (1) for prognostic purposes, it is necessary to know the parameter $k = 1/\sigma$ = const. If we use law (4) for prediction, then we need to know two parameters, *k* and *p*. In contrast, formula (10), similar to formulas (1) and (4), is not suitable for predictions, since information about the deactivation coefficient at some stage of evolution does not provide a basis for judging the subsequent behavior of the deactivation coefficient. Let us explain in this connection the meaning of elementary theory. It consists in the fact that, without making any a priori assumptions about the form of the function $\sigma(t)$, we calculate this function based on the aftershock observation data, using the solution (11) of the inverse source problem. In other words, the task of elementary theory is not to predict the activity of aftershocks, but to diagnose the relaxation process of the source.

The results of the experimental study of the source using the new method are presented in our review articles [Guglielmi, 2017; Zavyalov et al., 2020, 2022; Guglielmi et. al., 2023]. Here we will focus on a result that we consider to be quite significant. We were able to experimentally discover a two-stage relaxation mode of the source. At the first stage, called the Omori epoch, the criterion of applicability of the Omori law $\sigma$=const is strictly met. During the Omori epoch, the evolution of aftershocks is predictable. In the second stage, the deactivation coefficient changes unpredictably over time. Evolution at the second station is not described by either the Omori law or the Hirano-Utsu law. Function $\sigma(t)$ behaves non-monotonically, sometimes increasing, sometimes decreasing over time. The sharp transition from the first stage to the second resembles the phenomenon of bifurcation of a dynamic system.

Let us consider a specific example shown in Figure 5. The figure is based on data from the Northern California Regional Earthquake Catalog (http://www.ncedc.org) for 1968-2007. The zero moment in time coincides with the moment of the main shock. Let's use this event to present our research method in action.

The relaxation study of the source is carried out as follows. First, a discrete series of values of the auxiliary quantity $g = 1/n$ is compiled. It is shown in Figure 6. Then a spline approximation of the experimental points is performed and the deactivation coefficient $\sigma(t)$ is calculated using formula (11).

Function $\sigma(t)$ is shown in Figure 7. Visual analysis shows that the Omori epoch lasts approximately 20 days. The yellow bar in the figure indicates the time interval during which the condition $\sigma$=const is fulfilled with an accuracy of 1%. At the end of the epoch, something like a bifurcation occurs and the relaxation mode of the source changes dramatically.

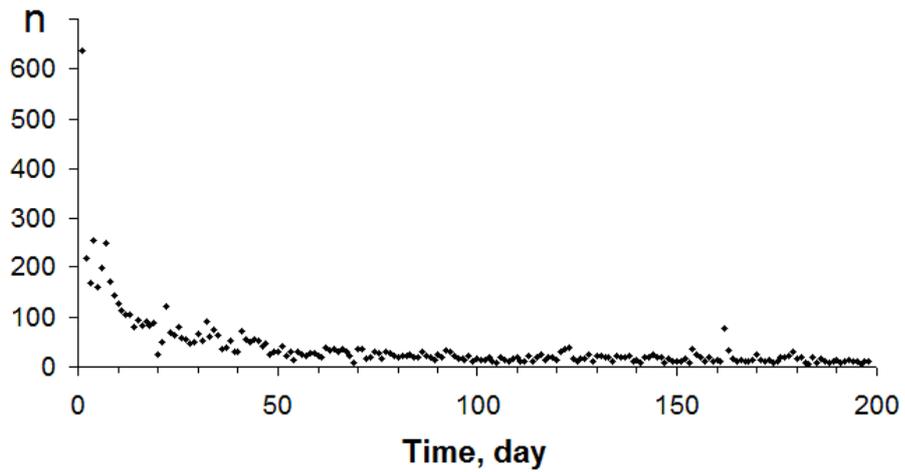

**Fig. 5**. Aftershock flow following the $M = 6$ earthquake that occurred at a depth of 10 km in Northern California on 11/23/1984. The vertical axis shows the number of aftershocks per day.

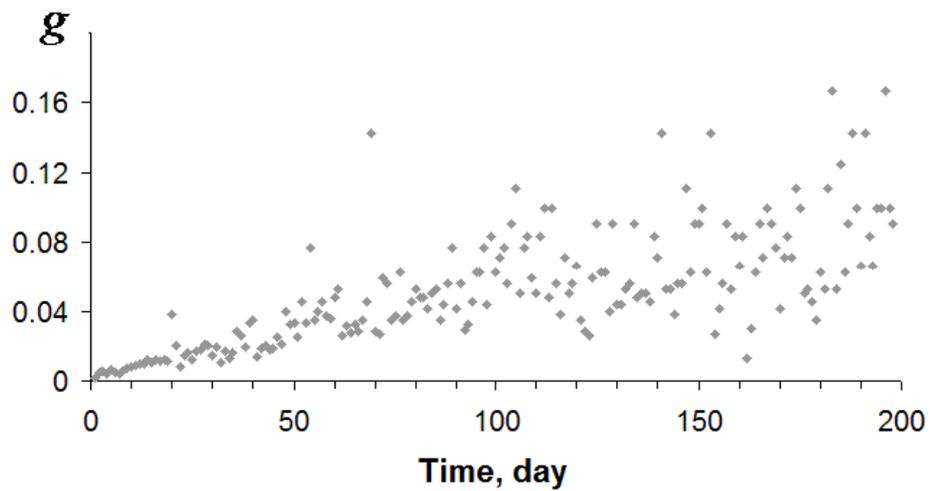

**Fig. 6**. Change over time of the auxiliary quantity $g = 1/n$ for the event shown in Fig. 5.

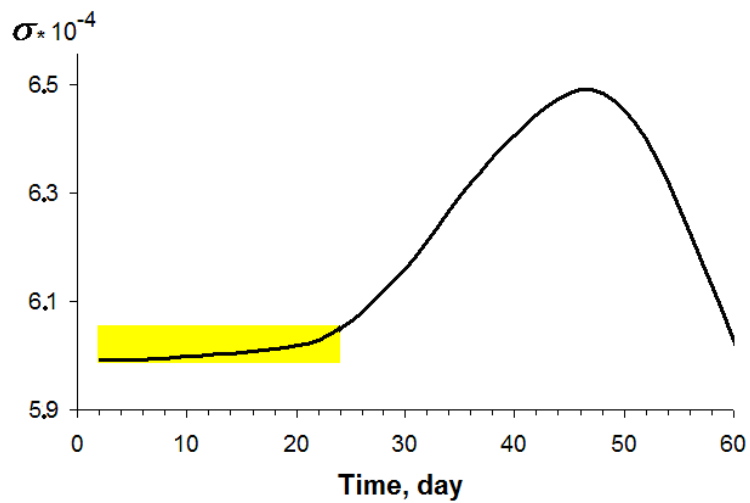

**Fig. 7**. The deactivation coefficient of the source as a function of time. The yellow bar indicates the Omori epoch.

The duration of the Omori epoch varies widely, from 10 to 100 days. There is a tendency for the duration to increase with increasing magnitude of the main shock.

The theory predicts a decrease in the deactivation coefficient with increasing magnitude of the main shock [Guglielmi, Zotov, 2021]. The experiment confirmed this prediction, as can be seen in Figure 8.

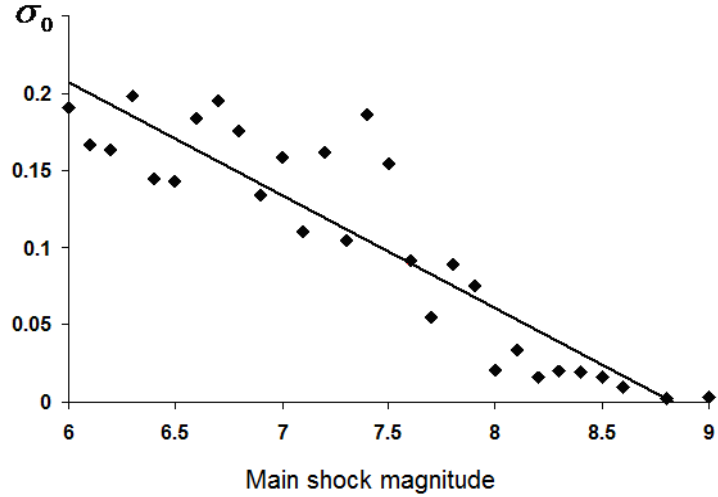

**Fig. 8**. Dependence of the deactivation coefficient in the Omori epoch on the magnitude of the main shock.

Thus, Omori's law (1) is stably fulfilled, but only at the first stage of relaxation of the source. In other words, Omori's law is not a holistic law, as its discoverer apparently implicitly assumed. The Hirano-Utsu law (4) is not fulfilled at either the first or the second stage of evolution. The only difference between the law of aftershock evolution (10) and the Omori law in its classical formulation (1) is that the world time $t$ is replaced by the so-called proper time of the earthquake source $\tau$.

What is the reason that the proper time of the source, as experience has shown, flows unevenly? The reason is as follows. We consider the source as a dynamic system whose properties are characterized by the parameter $\sigma$. It is quite obvious that our dynamic system is not closed. Accordingly, the parameter $\sigma$ changes over time under the influence of external influences, reflecting the non-stationarity of the rock mass that forms the source. We take into account the non-stationarity of the source by replacing $t$ with $\tau$.

## 4. Foreshocks
### 4.1. Gutenberg-Richter law

The mysterious variability of foreshock occurrence makes the prospect of monitoring foreshocks to predict mainshocks questionable. Nevertheless, foreshocks attract our attention as one of the signs of an impending catastrophe, or one of the "flags" in Gilmore's apt expression (see, for example, [Guglielmi, 2015]).

The Gutenberg-Richter law holds for foreshocks. The values of the key parameter $b$ in formula (2) for two classical, full symmetric and two types of mirror triads are given in Table 2. The initial data for compiling the table were selected according to the same rules that were indicated above in the description of Figure 4.

**Table 2.** Values of the parameter $b$ in the Gutenberg-Richter law for different types of triads.

| Specie of triad | | | | | |
|---|---|---|---|---|---|
| TC1 | TC2 | TS1 | TS2 | TM1 | TM2 |
| Foreshocks | | | | | |
| 1.03 | - | 1.28 | - | 1.41 | 1.7 |
| Main shock | | | | | |
| 0.4 | 0.5 | 0.8 | 1.3 | 0.7 | 1.0 |
| Aftershocks | | | | | |
| 1.19 | 1.31 | 1.81 | - | 1.73 | - |

The parameter $b$ in the Gutenberg-Richter law for foreshocks is one and a half to two times higher than for main shocks. Further, we see that the parameter $b$ in the complete classical triads is significantly smaller than in the small non-classical triads TS1, TM1 and TM2. Let us note that aftershocks have similar properties, as follows from examining the last row in Table 2. Taken together, the indicated ratios apparently indicate that the parameter $b$ reflects the properties of the rock mass that forms the source, which remain unchanged after the formation of the main rupture.

### 4.2. Mirror analogue of Bath's law

One of the authors (O.Z.) raised the question of finding a law for foreshocks similar to Bath's law for aftershocks, undertook a corresponding study and discovered the following.

For foreshocks in complete classical triads, a law similar to the classical Bath law established earlier for aftershocks holds true. Fig. 9 convincingly demonstrates what has been said. It is, in essence, a mirror image of Fig. 4. The difference concerns the quantitative values of the coefficients in the formula $\Delta M = aM_0 + b$. For foreshocks we have $a = 0.67$ and $b = -2.94$. The key parameter $a$ for foreshocks is one and a half times higher than for aftershocks. The geodynamic significance of this difference remains to be understood, but in general this property is associated with the fact that large earthquakes quite often do not have foreshocks. Note that the dependence $\Delta M = aM_0 + b$ can also be traced for foreshocks that are part of complete symmetric triads, namely $a = 0.47$ and $b = -2.0$. n this case, the parameter $a$ for foreshocks and aftershocks is almost the same, which is natural for symmetrical triads.

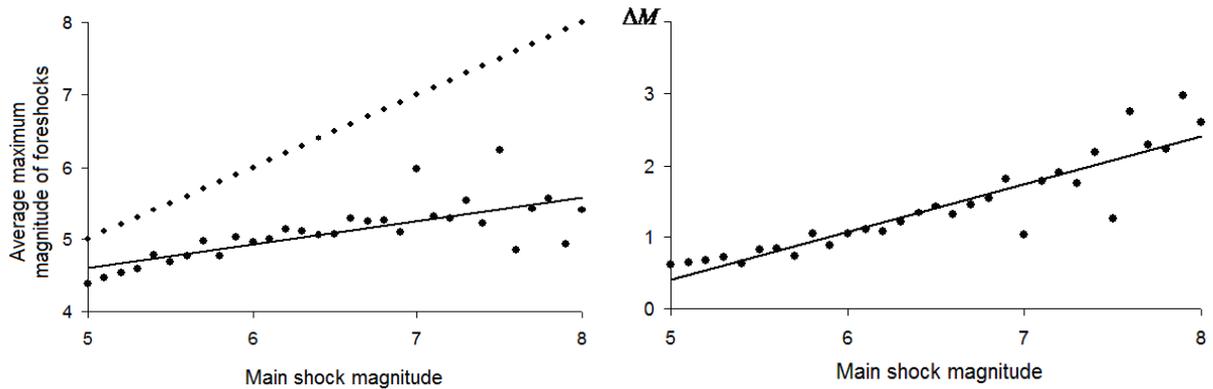

**Fig. 9**. Zotov's law, or the mirror analogue of Bath's law. The designations are the same as in Figure 4, with the significant caveat that they refer not to aftershocks, but to foreshocks of complete classical triads.

## 5. Discussion

The three laws that we have considered above, while not being postulates, i.e. self-evident statements, are recognized by the scientific community precisely as laws, i.e. reliably verified objectively existing stable relationships and connections between the properties of earthquakes. In contrast, the new propositions that emerged from our research and are briefly presented in this pfper do not have the legitimate status of laws. They have not undergone mandatory independent verification and therefore can only be considered as assumptions (hypotheses). However, in our opinion, they deserve attention.

The hyperbolic decrease in the aftershock frequency during the Omori epoch is abruptly terminated by a bifurcation. We have thus delimited Omori's law by indicating the limits of its applicability. However, our hypothesis requires independent experimental verification. The mirror analogue of Bath's law also needs to be verified and clarified.

The classification of main shocks as part of earthquake triads opens up an additional opportunity for detailing the geodynamic and tectonophysical conditions for the formation of cause-and-effect relationships leading to a main rupture in the continuity of rocks and determining the evolution of the source, "cooling" after the main shock.

We have attempted to present the elementary theory of aftershocks axiomatically. Taking the concept of source deactivation as a basis, we derived the equation for the evolution of aftershocks (9), which is convenient for processing and analyzing observation data. The analysis revealed a two-stage relaxation regime of the source after the main shock of an earthquake, with the Omori law in its classical formulation being fulfilled at the first stage.

The proposed theory is phenomenologically self-sufficient in the following sense. The state of the source as a dynamic system is described by a single phenomenological parameter – the deactivation coefficient. It is currently not known how to calculate the deactivation coefficient based on fundamental physical laws. However, as in other similar situations, the deactivation factor can be measured experimentally from observations of any particular aftershock sequence. Experience has shown that the deactivation coefficient changes over time, thereby generally reflecting the non-stationarity of the state of rocks in the source.

It must be said that the stable and objectively recognized patterns of earthquakes are not limited to the three laws that we selected for analysis.

Let us point out, for example, the connection between the magnitude of the main shock and the size of the earthquake source, which is important for seismology (see, for example, [Zavyalov, Zotov, 2021] and the literature cited in the paper). Wide opportunities for enriching the physics of earthquakes with new laws open up if we take into account that the source is actually a spatially extended object. An analysis of this problem would take us far from the topic of this paper. But one property of the spatial distribution of foreshocks and aftershocks deserves mention. We mean the divergence of aftershocks from the epicenter and the convergence of foreshocks to the epicenter of the main shock [Zotov et al., 2020; Guglielmi et al., 2023; Guglielmi, Zotov, 2024].

The laws of earthquake physics relate to average parameters and, quite naturally, are violated by disturbances of various origins. A special group of disturbances consists of endogenous and exogenous triggers that induce tremors. Let us indicate here two endogenous triggers generated by the main shock [Guglielmi et al., 2014]. One of them is pulsed and is a round-the-world seismic echo that originates when a main rupture is formed. The second, periodic trigger, is the free spheroidal and toroidal oscillations of the Earth, excited by the main shock.

In conclusion of the discussion, let us mention exogenous triggers of cosmic origin. Judging by the extensive literature related to this area of earthquake research, there is a certain connection between seismic activity and sporadic disturbances of the geomagnetic field (see, for example, [Guglielmi, Zotov, 2012; Zotov et al., 2013; Buchachenko, 2014]). Finally, we will mention the anthropogenic periodic impact on global seismic activity. In the work [Zotov, 2007], the so-called weekend effect in global seismic activity was discovered (see also [Sidorin, 2014; Ruzhin et al., 2016]).

## 6. Conclusion

Our synoptic review gives grounds to confidently assert that the laws of Omori, Gutenberg-Richter and Bath are a reliable support in the experimental and theoretical study of earthquakes. Expressing in a brief verbal and mathematical form objectively existing stable connections and relationships, the laws possess a depth of content, which manifested itself in the possibility of enriching the original formulations of the discoverers with interesting and important additional statements. So, Omori's law is strictly limited in time. Its action ends with a bifurcation of the source. The Gutenberg-Richter law demonstrates differences in a key parameter in the distribution of earthquakes belonging to different types of triads. Bath's law served as a model for us in searching for a mirror analog indicating an upper limit on the maximum magnitude of foreshocks. A careful study of the laws allowed us to see the problem of the earthquake source in a new light. It is significant that the study of deviations from laws not only strengthens our confidence in the statistical stability of fundamental statements, but serves as an incentive to search for new, previously unknown phenomena. This is how the search for the cumulative effect of the round-the-world seismic echo and the modulation of seismicity by free oscillations of the Earth was stimulated [Guglielmi et al., 2014; Zotov et al., 2018]. Our critique of the Hirano-Utsu law led us to the concept of the proper time of non-stationary lithospheric processes [Guglielmi et al., 2023], guided by which we discovered the convergence of foreshocks and the divergence of aftershocks.

The history of science shows that laws are not eternal. At one time, the geocentric system had the status of an immutable law of nature for a long time. Changes in our understanding of natural processes are natural and inevitable. But at this stage of the development of seismology, three fundamental laws, having stood the test of time, firmly form the basis of the physics of earthquakes.

*Acknowledgments*. We express our sincere gratitude to F.Z. Feygin and A.S. Potapov for support. We thank colleagues at the US Geological Survey for lending us their earthquake catalogs USGS/NEIC for use. The work was carried out within the framework of the planned tasks of the Ministry of Science and Higher Education of the Russian Federation to the Institute of Physics of the Earth of the Russian Academy of Sciences.